\documentclass[PAPER, cernpreprint=true, atlasdraft=false, USenglish, orcidlogo, texmf, orcidlogo]{atlasdoc}
\usepackage[subfigure,rotating]{atlaspackage}
\usepackage{atlasbiblatex}
\usepackage{makecell}
\usepackage{atlasphysics}
\graphicspath{{logos/}{figures/}}

\usepackage{ANA-STDM-2022-18-PAPER-defs}

\AtlasTitle{Observation of double parton scattering in same-sign W boson pair production in pp collisions at $\sqrt{s}=13$ TeV  with the ATLAS detector}

\AtlasAbstract{This letter reports the measurement of double parton scattering in same-sign W boson pair production with the ATLAS detector. The data set used corresponds to an integrated luminosity of 140~fb$^{-1}$ of proton--proton collisions at a center-of-mass energy of 13 TeV, collected during Run 2 of the Large Hadron Collider. The study is performed in final states including two same-charge leptons, electron or muon, missing transverse momentum, and up to one jet. An excess of events is observed over the expected background contributions with a significance of 8.8 standard deviations. The measured fiducial cross section times leptonic branching fraction is $4.59 \pm 0.64$ fb. The measurement corresponds to a double parton scattering effective cross section of $10.6 \pm 1.8$ mb.}

\AtlasRefCode{STDM-2022-18}

\PreprintIdNumber{CERN-EP-2025-109}

\AtlasJournalRef{Phys.\ Lett.\ B 870 (2025) 139892}
\AtlasDOI{10.1016/j.physletb.2025.139892}

\hypersetup{pdftitle={ATLAS document},pdfauthor={The ATLAS Collaboration}}

\begin{document}

\maketitle

\tableofcontents

\section{Introduction}

Multi-parton interactions (MPI) involving hard interactions of more than one pair of incident partons in the same proton--proton ($\pp$) collision have been discussed in theoretical studies since the first days of the parton model~\cite{PhysRevD.18.3344, PhysRevLett.43.1296, PhysRevD.22.2789}. This was followed by the generalization of the Altarelli–Parisi evolution equations to the case of multi-parton states~\cite{Kirschner:1979im, Shelest:1982dg} and theoretical studies of potential correlations in color and spin space~\cite{Mekhfi:1985dv}. Phenomenological studies of MPI in the framework of perturbative quantum chromodynamics (QCD) for a variety of processes leading to final states such as four leptons, four jets, three jets plus a photon, or a leptonically decaying gauge boson accompanied by two jets are discussed in Refs.~\cite{Paver:1982yp,PAVER1984252,PhysRevD.32.2371,Humpert:1983pw, HUMPERT1984179, Humpert:1984ay, Ametller:1985tp, Halzen:1986ue, godbole}.

The cross sections of the double-parton scattering (DPS) processes, where two partons in each proton initiate two separate hard-scattering processes, can be estimated by using a factorized ansatz that neglects potential complex correlation effects~\cite{Humpert:1984ay, Ametller:1985tp}. For a DPS process in which a final state A + B is produced at a hadronic center-of-mass energy $\sqrt{s}$, this simplified formalism yields

\begin{equation}
\sigma_{\mathrm{AB}}^{\mathrm{DPS}} = \frac{1}{1+\delta_{\mathrm{AB}}} \frac{\sigma_{\mathrm{A}} \sigma_{\mathrm{B}}}{\sigma_{\mathrm{eff}}},
\label{eq:DPS}
\end{equation}

where $\sigma_{\mathrm{AB}}^{\textrm{DPS}}$ denotes the DPS cross section and $\sigma_{\mathrm{A}}$ and $\sigma_{\mathrm{B}}$ denote the production cross sections of final states A and B in a single parton scattering (SPS), respectively. The quantity $\delta_{\mathrm{AB}}$ is the Kronecker delta used to construct a symmetry factor such that for identical final states with identical phase space, the DPS cross section is divided by two. The parameter $\sigma_{\mathrm{eff}}$, known as the effective cross section, describes the effective overlap of the spatial distribution of partons in the plane perpendicular to the direction of motion.  Experimental measurements of DPS effects in hadron collisions at different center-of-mass energies from 63~\GeV to 8~\TeV yield typical values of $\sigma_{\mathrm{eff}}$ ranging from about 10 mb at the lowest energy to 25 mb~\cite{STDM-2011-29,CMS-FSQ-12-028,sigeff4,CMS-GEN-14-001,sigeff3,BPHY-2014-01,Aaij:2015wpa,Abazov:2015fbl,CMS-BPH-21-004,sigeff5,sigeff6,cdf:dps,CDF:dps2,LHCb:dps,CMS-FSQ-16-005,STDM-2017-15,CMS-SMP-20-007,CMS-SMP-20-009,D0:2014vql,D0:2014owy, D0:2015rpo, STDM-2015-08, BPHY-2015-02, CMS-BPH-14-008, LHCb:2016wuo}.

The study of DPS events is crucial for fully understanding the internal structure of the colliding hadrons, including the correlation effects among the partons that are neglected in the simplified approach of Eq.~\ref{eq:DPS}. Recent theoretical developments~\cite{Gaunt:2009re,Diehl:2017kgu,Diehl:2018kgr,Gaunt:2018eix,Cabouat:2019gtm} have introduced non-factorized double parton distribution functions including perturbative splittings with impact parameter dependence. The high energy and high integrated luminosity available at the Large Hadron Collider~\cite{Evans:2008zzb} (LHC) offers the possibility to further study these interactions in a variety of processes.

\begin{figure*}[t!bp]
\centering
\subfigure[]{\includegraphics[width=.32\textwidth]{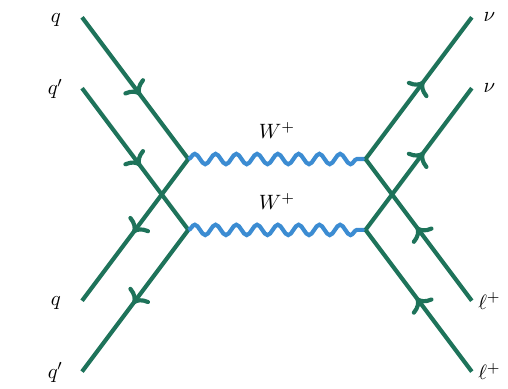}}
\subfigure[]{\includegraphics[width=.30\textwidth]{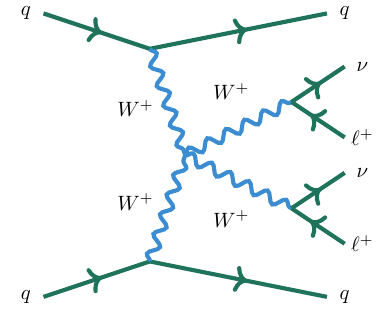}}
\caption{
Representative Feynman diagrams of $W^+W^+$ process produced via (a) DPS and (b) SPS.}
\label{fig:feynman}
\end{figure*}

This letter presents the measurement of DPS production of same-sign $W$ boson pairs (\ssww) with the ATLAS detector. Observation of the DPS \ssww production at 13~\TeV with a significance greater than 5 standard deviations was reported by the CMS Collaboration~\cite{CMS-SMP-21-013}. The \ssww process is a promising channel to study DPS~\cite{Gaunt:2010pi,Ceccopieri:2017oqe}  compared to opposite-sign $W^{\pm}W^{\mp}$ because of the smaller background contribution from the production of SPS \ssww. The SPS \ssww process can indeed be suppressed by selections as two additional partons are needed in the final state at the leading-order (LO) accuracy to produce same-sign W bosons, as seen in Figure~\ref{fig:feynman}.

The measurement is performed in the leptonic decay modes $\ssww \to \ell^\pm\nu\ell'^\pm\nu$, where both $W$ bosons decay into electrons or muons, $\ell, \ell' = \mathrm{e}$, $\mu$. A set of $\pp$ collision data events collected by the ATLAS detector between 2015 and 2018 at $\sqrt{s}$ = 13~\TeV is used. The dataset corresponds to an integrated luminosity of 140 fb$^{-1}$~\cite{DAPR-2021-01,LUCID2} with a relative uncertainty of 0.83\%.  Candidate events contain two same charge leptons, moderate missing transverse momentum, and at most one jet. As discussed above, the contribution of the SPS \ssww background process is not significant after the requirement on the number of jets. The leading Standard Model (SM) background process is the production of $WZ/\gamma^*$ boson pairs (referred to as $WZ$). It contributes when the lepton from the $Z$ boson decay, having an opposite sign to that of the $W$ boson, is not detected, typically because it is outside of the geometrical and kinematical acceptances of the detector. Its contribution is estimated using Monte Carlo (MC) simulated events, with the normalization constrained in a dedicated signal-depleted control region (CR).  Data-driven techniques, assisted by MC simulation, are used to estimate backgrounds including electrons or muons not originating from the prompt decay of particles such as $W$ or $Z$ bosons (referred to as non-prompt leptons) and backgrounds including electron charge misidentification. Background events from the production of two overlapping same-sign $W$ bosons originating from separate  $pp$ interactions within the same bunch crossing are also considered. Other smaller backgrounds, including contributions mainly from the $V\gamma$~($V=W,Z/\gamma^*$), SPS \ssww, $ZZ$, and top quark processes, are estimated using MC simulation.

\section{ATLAS detector}

The ATLAS experiment~\cite{PERF-2007-01} at the LHC is a multipurpose particle detector
with a forward--backward symmetric cylindrical geometry and a near \(4\pi\) coverage in
solid angle.\footnote{ATLAS uses a right-handed coordinate system with its origin at the nominal interaction point (IP) in the center of the detector and the \(z\)-axis along the beam pipe. The \(x\)-axis points from the IP to the center of the LHC ring, and the \(y\)-axis points upwards. Polar coordinates \((r,\phi)\) are used in the transverse plane, \(\phi\) being the azimuthal angle around the \(z\)-axis. The pseudorapidity is defined in terms of the polar angle \(\theta\) as \(\eta = -\ln \tan(\theta/2)\) and is equal to the rapidity $ y = \frac{1}{2} \ln \left( \frac{E + p_z}{E - p_z} \right) $ in the relativistic limit. Transverse momentum ($\pt$) is defined relative to the beam axis and is calculated as $p_T = p \sin \theta$ where $p$ is the momentum. Angular distance is measured in units of \(\Delta R \equiv \sqrt{(\Delta y)^{2} + (\Delta\phi)^{2}}\).}
It consists of an inner tracking detector (ID) surrounded by a thin superconducting solenoid
providing a \qty{2}{\tesla} axial magnetic field, electromagnetic and hadronic calorimeters, and a muon spectrometer.
The ID covers the pseudorapidity range \(|\eta| < 2.5\).
It consists of silicon pixel, silicon microstrip, and transition radiation tracking detectors.
Lead/liquid-argon (LAr) sampling calorimeters provide electromagnetic (EM) energy measurements
with high granularity within the region \(|\eta|< 3.2\).
A steel/scintillator-tile hadronic calorimeter covers the central pseudorapidity range (\(|\eta| < 1.7\)).
The endcap and forward regions are instrumented with LAr calorimeters
for EM and hadronic energy measurements up to \(|\eta| = 4.9\).
The muon spectrometer (MS) surrounds the calorimeters and is based on
three large superconducting air-core toroidal magnets with eight coils each.
The MS includes a system of precision tracking chambers up to \(|\eta| = 2.7\) and fast detectors for triggering up to \(|\eta| = 2.4\).
The luminosity is measured mainly by the LUCID--2~\cite{LUCID2} detector, which is located close to the beampipe.
A two-level trigger system is used to select events~\cite{TRIG-2016-01}.
The first-level trigger is implemented in hardware and uses a subset of the detector information
to accept events at a rate close to \qty{100}{\kHz}.
This is followed by a software-based trigger that
reduces the accepted rate of complete events to \qty{1.25}{\kHz} on average
depending on the data-taking conditions.
A software suite~\cite{SOFT-2022-02} is used in data simulation, in the reconstruction
and analysis of real and simulated data, in detector operations, and in the trigger and data acquisition
systems of the experiment.

\section{Event simulation}

MC simulated event samples are used to model the signal DPS \ssww process as well as other background processes. Simulated events are processed through the ATLAS simulation infrastructure~\cite{SOFT-2010-01} using \textsc{Geant4}~\cite{Agostinelli:2002hh}. The effect of additional $\pp$ interactions per bunch crossing (\pileup) is accounted for by overlaying the hard-scattering process with Poisson-distributed minimum-bias events generated with \PYTHIA[8.186]~\cite{Sjostrand:2007gs} using the \NNPDF[2.3lo] set of parton distribution functions (PDF)~\cite{Ball:2012cx} and the A3 set of tuned parameters~\cite{ATL-PHYS-PUB-2016-017}. Different \pileup conditions between data and simulation are taken into account by reweighting the mean number of interactions per bunch crossing in simulation to the number observed in data. The \EVTGEN[1.7.0] program~\cite{Lange:2001uf} was used to model the decays of bottom and charm hadrons. All simulated samples are processed through the same reconstruction algorithms and analysis chain as the data.

The DPS \ssww production is simulated at LO using the \PYTHIA[8.307]~\cite{Bierlich:2022pfr} generator with the A14~\cite{ATL-PHYS-PUB-2014-021} set of tuned parameters used for the parton shower. The \NNPDF[2.3lo] PDF set is used. An alternative signal sample is simulated at LO with the \HERWIG[7.2.13]~\cite{Bahr:2008pv,Bellm:2015jjp} generator using the default \HERWIG tune and the \NNPDF[3.0lo] PDF set~\cite{Ball:2014uwa}. Differences in the shapes of the signal distributions predicted by the \PYTHIA and \HERWIG samples are considered as an uncertainty.

Detailed information about the simulation of the background processes can be found in Ref.~\cite{STDM-2018-32}. A brief description is given below. The \NNPDF[3.0]~\cite{Ball:2014uwa} PDF sets are used in all matrix element calculations. The $WZ$ process, where both the $W$ and $Z$ bosons decay leptonically, is simulated with the \SHERPA[2.2.12] generator~\cite{Bothmann:2019yzt} using matrix elements that contain all diagrams with four electroweak (EW) vertices. This process is calculated for up to one additional parton at next-to-LO (NLO) in perturbative QCD and up to three additional partons at LO using \textsc{Comix}~\cite{Gleisberg:2008fv} and \OPENLOOPS~\cite{Cascioli:2011va}, and merged with the \SHERPA parton shower based on the Catani--Seymour dipole factorization~\cite{Schumann:2007mg}. The $WZ$ production in association with two jets involving only  EW vertices as well as the interference between the EW and strong contributions are simulated using \MGNLO{2.6.2}~\cite{Alwall:2014hca} at LO interfaced to \PYTHIA[8.235]~\cite{Sjostrand:2014zea} for modeling the parton shower in the dipole recoil scheme~\cite{Cabouat:2017rzi}.

The production of the SPS $\ssww$ processes at LO has contributions both from processes that involve only EW interaction vertices, referred to as EW \ssww and from processes that involve strong interaction vertices, referred to as QCD \ssww. The SPS \ssww contributions are simulated with \MGNLO[2.6.7] at LO. The samples corresponding to the EW \ssww production and decay are simulated with diagrams including exactly six orders of the EW coupling~\cite{STDM-2018-32}. The simulation of the QCD \ssww process and the interference between EW and QCD contributions includes diagrams with exactly four and five EW vertices, respectively.

The diboson $ZZ$ production is simulated using the \SHERPA[2.2.2] generator. The $W\gamma$ and $Z\gamma$ processes are simulated with \SHERPA[2.2.11] with all off-shell contributions included. The NLO matrix elements with up to one additional parton and LO matrix elements with up to three partons are merged with the parton shower using an MEPS@NLO merging scale~\cite{Hoeche:2011fd,Hoeche:2012yf,Catani:2001cc,Hoeche:2009rj} of $Q=20$~\GeV.

Additional samples are used to model minor backgrounds. The production of \ttV and \tZq events are simulated at NLO and LO in QCD, respectively, using the \MGNLO[2.3.3] generator interfaced to \PYTHIA[8.210] for parton showering. The production of $t\bar{t}$ and single-top-quark events is simulated using the \POWHEGBOX[v2]~\cite{Frederix:2012dh,Nason:2004rx,Frixione:2007vw,Alioli:2010xd}
generator at NLO in QCD and the production of $V+$jets events~\cite{PMGR-2021-01} is simulated with the \MGNLO[2.3.2] generator. \PYTHIA[8.230] with the A14 set of tuned parameters is used for the parton shower. The production of triboson ($VVV$) events is simulated with the \SHERPA[2.2.2] generator, accurate at NLO in QCD for the inclusive process and at LO for up to two additional parton emissions, using factorized gauge-boson decays. These background samples are normalized to the highest-order theory predictions available as described in Ref.~\cite{STDM-2018-32}.

\section{Object reconstruction and event selection}

Events are selected online by a set of single-electron~\cite{TRIG-2018-05} or single-muon triggers~\cite{TRIG-2018-01}, with lepton \pT lowest thresholds ranging from 20 to 26~\GeV, depending on the lepton flavor and data-taking period. Only data taken from stable beam collisions that satisfy a standard set of data-quality requirements, ensuring that all the ATLAS subdetectors were functioning correctly, are considered~\cite{DAPR-2018-01}. Events are required to have at least one collision vertex reconstructed from at least two ID tracks with \(\pt > 500~\MeV\). For events with several collision vertices, the one with the largest sum of the squared transverse momenta of the associated tracks is taken as the hard-scatter vertex~\cite{ATL-PHYS-PUB-2015-026}.

Electrons are reconstructed from isolated electromagnetic calorimeter clusters, which are matched to tracks in the ID~\cite{EGAM-2018-01}. ``Baseline" electrons are required to satisfy a likelihood-based identification criterion with $\pt >4.5~\GeV$ and $|\eta |<$ 2.47 with an average efficiency of 93\%, referred to as ``Loose" working point in Ref.~\cite{EGAM-2018-01}. The electrons must be outside the barrel/endcap transition region ($1.37 < |\eta| < 1.52$) of the calorimeter. The transverse impact parameter significance\footnote{The transverse impact parameter significance is defined as $|d_0|/\sigma(d_0)$, where $d_0$ is the distance of closest approach of the $e$ or $\mu$ track to the \(z\)-axis in the transverse plane and $\sigma(d_0)$ is its uncertainty.} is required to satisfy ${|\dzero|/\sigma(\dzero)} < 5$.  The longitudinal impact parameter\footnote{The longitudinal impact parameter is equal to $|z_0 \sin \theta|$, where $z_0$ is the difference between the value of the $z$ coordinate of the point on the track at which $d_0$ is defined, and the longitudinal position of the primary vertex.} is required to satisfy ${|\zzsth|<0.5}$~mm. ``Signal" electrons have the same requirements as baseline electrons but with $\pt>27~\GeV$ and must additionally satisfy a more restrictive (``Tight'') likelihood-based identification with an average efficiency of 80\% and ``Gradient" isolation requirements~\cite{EGAM-2018-01}. A charge selector tool based on boosted decision trees uses shower shape and track-to-cluster matching variables~\cite{EGAM-2018-01} to reject electron candidates where the charge is likely misidentified. \enquote{Background} electrons, used to estimate the background processes with non-prompt electrons, are required to pass ``Medium" likelihood-based identification with an average efficiency of 88\%~\cite{EGAM-2018-01}, with no requirements on the isolation criteria. Background electrons are required to fail the signal electron selection criteria to ensure that the samples of signal and background electrons are statistically independent.

Muons are reconstructed~\cite{MUON-2018-03} from tracks in the MS, matched to a corresponding track in the ID where possible. Baseline muons are required to have $\pT>3~\GeV$ and $|\eta|<2.7$ with ${|\dzero|/\sigma(\dzero)} < 15$ and ${|\zzsth|<1.5}$~mm.  Baseline muon candidates must satisfy the ``Loose" cut-based identification criteria with an average efficiency of 99\%~\cite{MUON-2018-03}. Signal muons must satisfy a more restrictive identification criteria (“Medium") with an average efficiency of 97\% with the ``PflowTight" isolation requirement~\cite{MUON-2018-03}. The signal muons must have $\pt>27~\GeV$ and are restricted to the range $|\eta| < 2.5$ with ${|\dzero|/\sigma(\dzero)} < 3$ and ${|\zzsth|<0.5}$~mm. The isolation requirement for background muons, used to estimate the background processes with non-prompt muons, is changed to a less restrictive working point (`PflowLoose'), with an efficiency of 99\%, with ${|\dzero|/\sigma(\dzero)} < 10$~\cite{MUON-2018-03}. Background muons are required to fail the signal muon selection to ensure that the samples of signal and background muons are statistically independent. The muon charge misidentification rate is found to be negligible~\cite{STDM-2010-16}.

Jets are reconstructed using the \antikt algorithm~\cite{Cacciari:2008gp,Fastjet}, with a radius parameter of \(R=0.4\), using particle-flow objects~\cite{PERF-2015-09} as inputs. Contamination from jets originating from \pileup is reduced by using the jet-vertex-tagger algorithm~\cite{PERF-2014-03}. The jets are calibrated as described in Ref.~\cite{JETM-2018-05} and required to have $\pt > 30~\GeV$ and $|\eta| \le 4.5$. In order to suppress contributions from background processes that involve top quarks or leptonic $b$-hadron decays, the DL1r classification algorithm based on recurrent neural networks~\cite{FTAG-2019-07} is used to identify jets originating from  $b$-quarks, referred to as \enquote{\(b\)-jets}.  The \(b\)-jets with $\pT > 20~\GeV$ and $|\eta| < 2.5$ have an identification efficiency of \SI{85}{\percent} in $\ttbar$ events with an expected rejection factor (defined as the inverse of the efficiency) of about 40 for light-flavor jets, and about 2.9 for jets originating from charm quarks~\cite{FTAG-2018-01, FTAG-2019-02, FTAG-2020-08}.

An object overlap removal procedure is applied to baseline leptons and jets to avoid ambiguities in detector reconstruction. This procedure prevents cases where the detector response to a single physical object is reconstructed as two different final state objects. First, this procedure removes any electron if it shares an ID track with another higher $\pt$ electron. Second, electrons sharing their tracks with a muon candidate are removed. Third, a jet is removed if it overlaps with an electron within a $\Delta R$ distance of 0.2, unless it is a $b$-jet and the electron \pt is below 100~\GeV, in which case the electron is removed, and any electrons within $\Delta R = 0.4$ of a remaining jet are removed. Any jet that is within $\Delta R = 0.2$ of a muon and has less than three associated tracks is removed, unless it is a $b$-jet. Lastly, the remaining muons are removed if their track is within $\Delta R = 0.4$ of a remaining jet.

The missing transverse momentum, with magnitude \MET, is calculated from the negative vector sum of the transverse momenta of all of the selected and calibrated objects in the event including a track-based soft term~\cite{JETM-2020-03}.

Candidate events in the signal region (SR) are required to have a same-sign signal lepton pair, with an invariant mass $m_{\ell\ell}$ greater than $20$~\GeV to reduce the non-prompt lepton background contribution. One of the leptons must be matched to the lepton that fired a single-lepton trigger. Events with an additional baseline muon or electron are vetoed to reduce the contributions from the $WZ$ and $ZZ$ background processes. Events with $b$-jets are discarded, reducing the non-prompt lepton background and signal contributions by 36\% and 4\%, respectively. The \MET must be larger than 30~\GeV to exploit the presence of neutrinos in the final state and to reduce the contribution of $Z/\gamma* \to ee$ events where the charge of the electron is misidentified. For the $ee$ final state, both electrons are required to be in the barrel region with $|\eta|<1.37$, to further reduce backgrounds with electron charge misidentification. In addition, if a third baseline lepton does not pass the overlap removal requirement, but the $m_{\ell\ell}$ of a same-flavor opposite-charge signal lepton and the third lepton is compatible with the $Z$ boson mass, $|m_{\ell\ell} - m_{Z}| < 15~\GeV$, the event is also rejected. Events with two or more selected jets are vetoed to suppress the contribution of the SPS \ssww background.

The normalization of the $WZ$ process is constrained in a dedicated signal-depleted CR defined closely following the SR selection but requiring three charged leptons in the final state, two of which have opposite charge in order to be compatible with a $Z$ boson decay. The third lepton is required to satisfy $\pt > 15~\GeV$. Events containing a fourth baseline electron or muon are removed to reject events from the $ZZ$ background process. The trilepton invariant mass is required to be greater than 106~\GeV.

\section{Analysis strategy}

The predicted DPS \ssww contribution changes significantly depending on the number of exclusive jets in the SR. Therefore, the events in the SR are categorized according to the number of jets in the final state, in exclusive $0-$jet and $1-$jet categories, where the expected signal contribution in the $0-$jet category is approximately $70\%$ of the overall expected signal yield. The composition of the non-prompt and charge misidentification background processes is lepton-\pT and flavor dependent. The SR in each jet category is further split into four categories depending on the flavors of the leading- and subleading-\pT leptons: \(ee\), \(e\mu\), \(\mu e\), \(\mu\mu\), to explore the different signal-to-background ratios and to enable better constraint on the background uncertainties.

\begin{figure}[t!bp]
\centering
\subfigure[]{\includegraphics[width=0.45\textwidth]{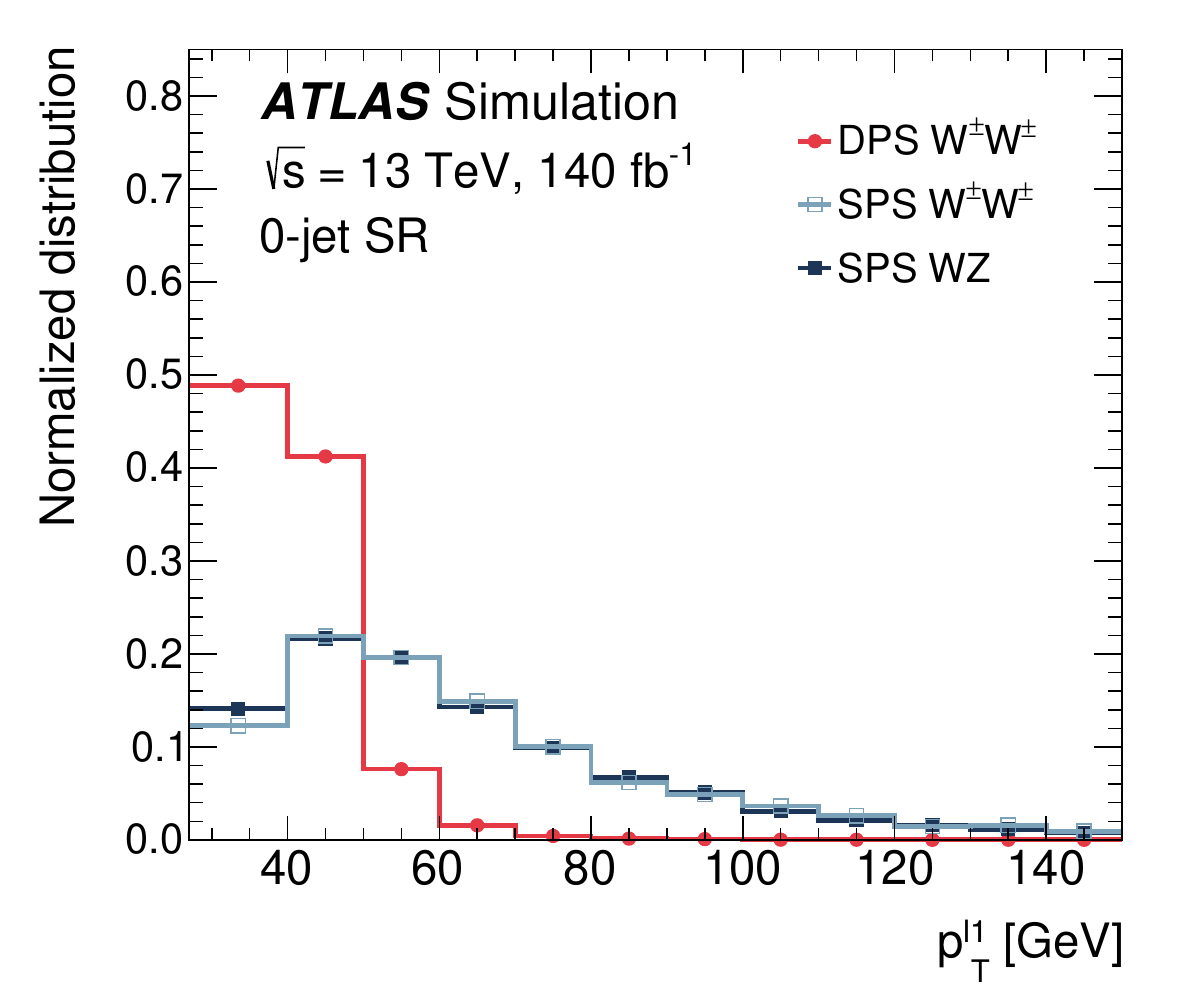}}
\subfigure[]{\includegraphics[width=0.45\textwidth]{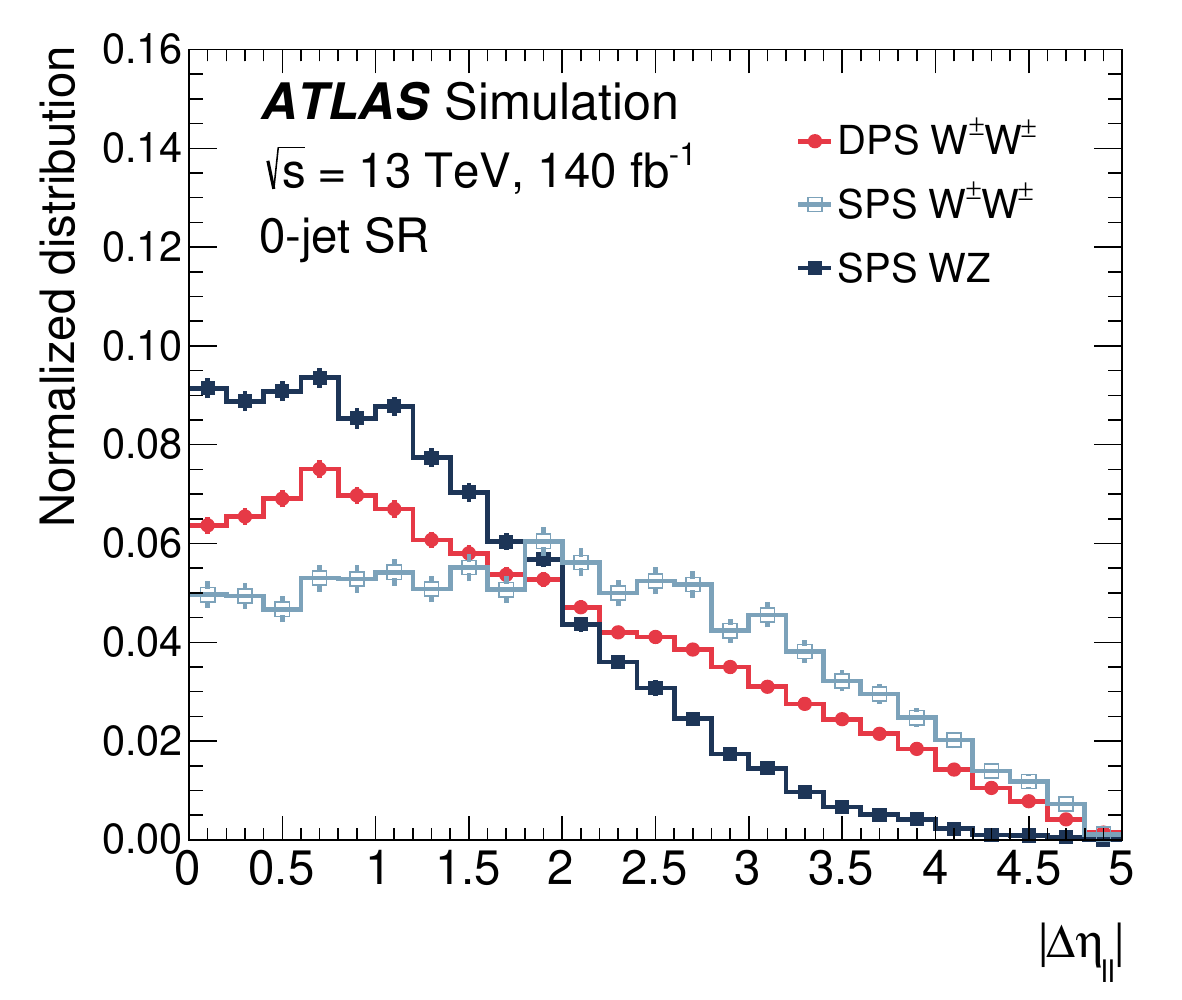}}
\subfigure[]{\includegraphics[width=0.45\textwidth]{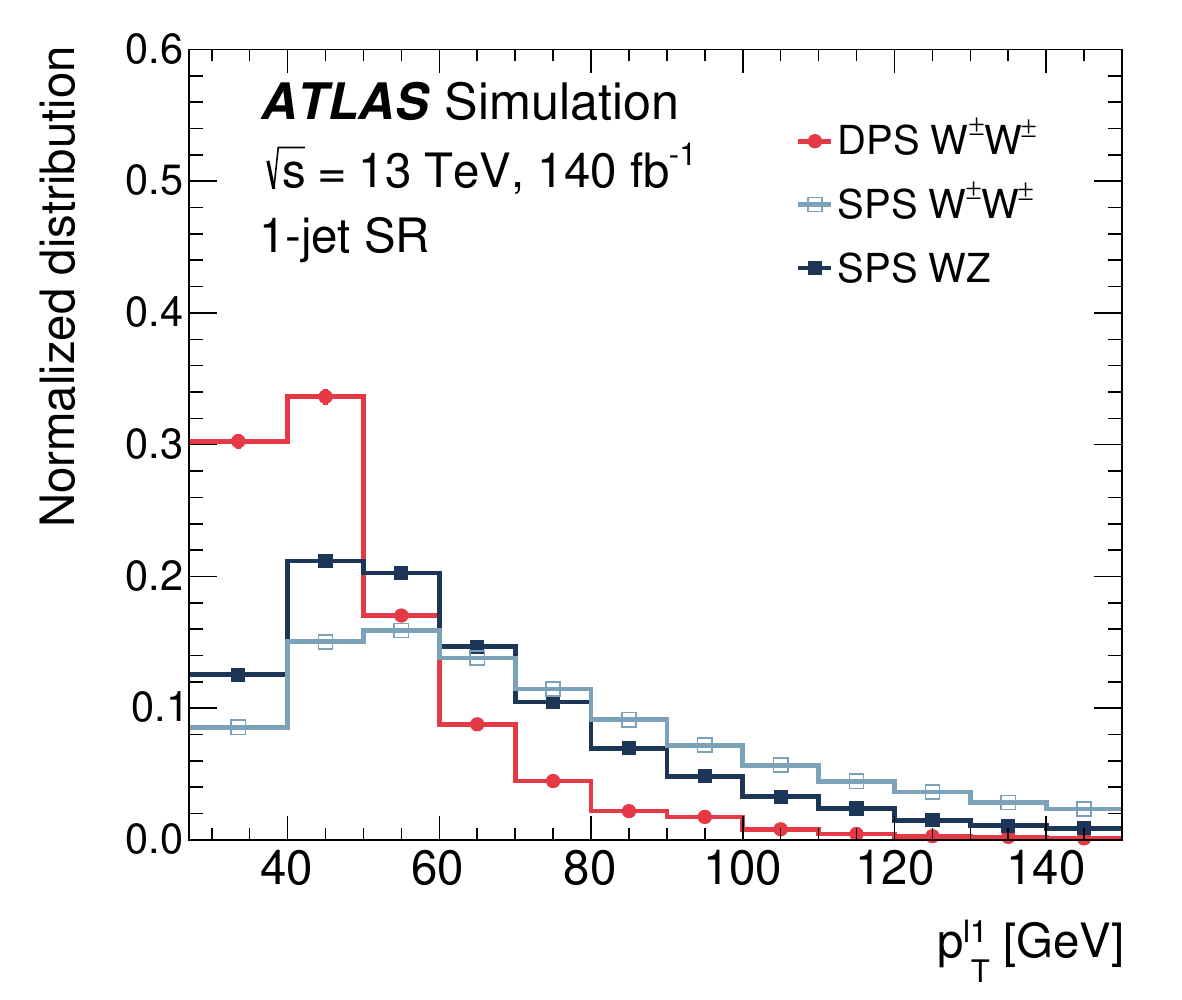}}
\subfigure[]{\includegraphics[width=0.45\textwidth]{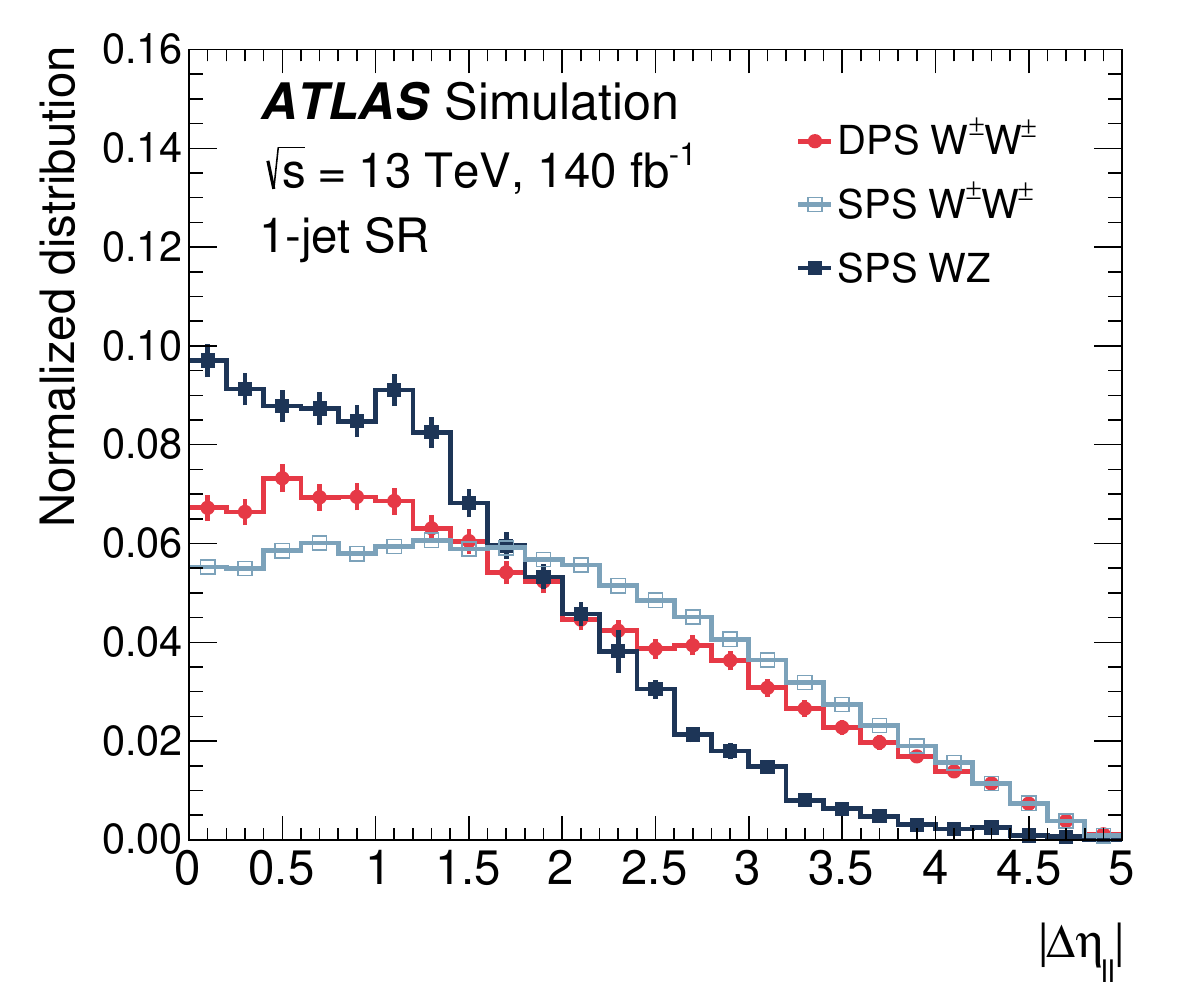}}
\caption{Distributions of (a and c) $\pt^{\ell 1}$ and (b and d) $|\Delta \eta_{\ell\ell}|$ in the SR (a and b) $0-$jet category and (c and d) $1-$jet category  for the DPS \ssww, SPS \ssww, and $WZ$ processes. The error bars represent the statistical uncertainties due to the limited number of simulated events.}
\label{fig:var_plots}
\end{figure}

To extract the DPS \ssww production, two independent deep neural networks (DNNs) are trained on MC simulated events in the $0-$jet and $1-$jet categories, aiming to separate the DPS signal from the dominant $WZ$ background process. Dedicated optimizations of the model structures and hyperparameters are performed for each DNN with 3 hidden layers and up to 127 neurons in the first layer. The kinematic variables showing the best discrimination, evaluated by removing each variable and retraining the network, are retained. The variable importance is also evaluated with the \textsc{SHAP} framework~\cite{NIPS2017_8a20a862}. Only lepton kinematic variables are considered to avoid potential biases in the modeling of jet-related variables, as the LO DPS \ssww MC simulation models jets via parton showering. Eight such kinematic variables are retained from a larger set as summarized in Table~\ref{tab:input_variables}. The $\pt^{\ell 1}$ and $|\Delta \eta_{\ell\ell}|$ variables provide the strongest discrimination for the signal DNNs. The simulated distributions of these variables in the SR $0-$jet and $1-$jet categories for the DPS \ssww, SPS \ssww, and $WZ$ processes are shown in Figure~\ref{fig:var_plots}.

\begin{table}[t!bp]
\caption{Lepton kinematic variables used in the training and optimization of the DNNs.}
\label{tab:input_variables}
\centering
\resizebox{\linewidth}{!}{
\begin{tabular}{p{2ex}llc} %
\toprule %
\multicolumn{2}{l}{Description} & Variable\\ %
\midrule %
& \pt of the leading lepton & $\pt^{\ell 1}$ \\ %
& \pt of the subleading lepton & $\pt^{\ell 2}$ \\ %
& Invariant mass of the dilepton system & $m_{\ell \ell}$ \\ %
& Pseudorapidity of the dilepton system & $\eta_{\ell \ell}$ \\ %
& Difference in azimuthal angle between the leading and subleading leptons & $\Delta \phi_{\ell\ell}$ \\ %
& Difference in pseudorapidity between the leading and subleading leptons  & $|\Delta \eta_{\ell\ell}|$ \\ %
& Distance $\Delta R$ between the leading and subleading leptons & $\Delta R_{\ell\ell}$ \\ %
& Difference in azimuthal angle between the missing transverse momentum and subleading lepton & $\Delta \phi_{\ell_2, \MET}$  \\ %
\bottomrule %
\end{tabular}} %
\end{table}

The non-prompt background contribution, where either a hadron or a lepton from a hadron decay satisfies the signal lepton selection criteria, is estimated using the data-driven fake-factor method~\cite{EGAM-2019-01}. Fake factors are determined as functions of electron or muon $\pT$ and $\eta$ in a dedicated region enriched in non-prompt leptons. This region is selected from collision data by requiring events in which jets recoil against a non-prompt lepton candidates~\cite{STDM-2018-32}. A back-to-back topology between the jet and lepton is enforced by imposing a minimum azimuthal angle separation between the lepton and the jet, $\Delta\phi_{\ell{j}} > 2.8$. Contributions from prompt leptons produced in $W$, $Z$, or top-quark decays, as well as from photon conversions in $\gamma$+jet events, are subtracted in this region using MC simulation. The region is designed to closely match the composition of non-prompt lepton sources present in the SR, ensuring the validity of the fake factors when estimating the background contribution in the SR. The fake factors for electrons and muons are evaluated independently as the ratio of events where the non-prompt lepton passes the signal lepton selection to those where it passes the background lepton selection. The final estimate of the non-prompt background in the SR is obtained by re-scaling data from the same SR event selection but with background leptons by the corresponding fake factors~\cite{EGAM-2019-01}.

Another significant background source arises from opposite-sign lepton pair events in which one of the two lepton charges has been misidentified. This effect is more prevalent for electrons due to bremsstrahlung radiation followed by electron--positron pair production. This background, referred to as \textit{charge flip}, is estimated by re-weighting opposite-sign events in data with a factor representing the probability of charge misidentification in electrons~\cite{EGAM-2018-01}.  The charge misidentification rate is determined as a function of electron \pt and $\eta$, using simulated \(Z \to e^+e^-\) events that are corrected to match the misidentification rate in data through dedicated scale factors~\cite{STDM-2018-32}. The uncertainty in this background contribution is 40\%, mainly driven by uncertainties in the scale factors used to calibrate the charge-selector tool, which are derived from \( Z \to e^+e^- \) events~\cite{EGAM-2018-01}.

The background from $\ell \gamma$ events, where the prompt photon \( \gamma \) is misidentified as an electron, is modeled using the \( W\gamma \) simulated sample, and is further suppressed by the electron charge-selector tool. This background is assigned an overall normalization uncertainty of 40\%, motivated by the scale factor uncertainties of the charge-selector tool, following the approach used in Ref.~\cite{STDM-2018-03}. The small contributions of prompt lepton backgrounds from the SPS \ssww, $ZZ$, $VVV$, and top quark processes are modeled using MC samples.

The contribution to the signal yield from the production of two overlapping same-sign $W$ bosons originating from separate \pp interactions within the same bunch crossing, referred to as pileup background, is largely reduced by the requirements on $|\zzsth|$. This background contribution is estimated using a toy study based on the average number of \pp interactions per bunch crossing, the longitudinal size of the beam spot, and the number of expected $W^{+}$ and $W^{-}$ events in the fiducial region, following the method described in Ref.~\cite{ATL-PHYS-PUB-2018-007}. The resulting contribution in the SR is estimated to be approximately 4.5\% of the total signal yield. A signal template normalized to this contribution is used to model this background, which constitutes less than 1\% of the total background in the SR. Further reduction of this background by an application of a more restrictive requirement on $|\zzsth|$ is not considered as the contribution is sufficiently small.

Systematic uncertainties in this measurement arise from experimental and theory sources. The results are driven by the statistical uncertainty of the data in the SR and none of the considered systematic uncertainties have a significant impact on the sensitivity of this result. Experimental systematic uncertainties are related to the trigger, lepton reconstruction, identification and isolation efficiencies~\cite{MUON-2018-03, EGAM-2018-01}, lepton energy (momentum) scale and resolution~\cite{MUON-2018-03, EGAM-2018-01}, jet energy scale and resolution~\cite{JETM-2018-05}, $b$-jet identification~\cite{FTAG-2018-01}, modeling of \MET~\cite{JETM-2020-03}, and integrated luminosity~\cite{DAPR-2021-01,LUCID2}. The uncertainty in the pileup reweighting procedure is considered by varying the average number of \pp collisions in simulated events to cover the uncertainty in the measured inelastic $\pp$ cross-section~\cite{STDM-2015-05}. The uncertainties related to the non-prompt lepton background is studied in detail. Three sources of systematic uncertainty are considered including the statistical error on the fake factors, uncertainties in the composition of the fake factor CR obtained by varying the $b$-jet requirement in this region, and uncertainties related to the prompt lepton contribution in the fake-factor CR~\cite{STDM-2018-32}. Uncertainties in the data-driven charge flip background are described in Ref.~\cite{EGAM-2018-01}. The theory uncertainties in the physics modeling of the background processes are estimated by varying the factorization and renormalization scales, the strong coupling constant $\alpha_S$, and the choice of the PDF~\cite{Butterworth:2015oua}. The theory uncertainties in the modeling of the DPS \ssww signal include scale variations and an uncertainty related to the difference in the shapes of the signal distributions predicted by the \PYTHIA and \HERWIG samples. The systematic uncertainties with the largest impact on the result are related to the jet and \MET scale and resolution and the non-prompt lepton background estimation.

\section{Results}

A binned maximum-likelihood fit is performed for the DPS \ssww process using the DNN score distributions in the $0-$jet and $1-$jet categories. The DNN score boundaries are chosen independently in the two exclusive jet categories to maximize the expected significance of the DPS \ssww process. The total number of events in the $WZ$ CR and the four dilepton final states in the SR are fitted simultaneously in the two exclusive jet categories with the DPS \ssww normalization and the two normalizations of the $WZ$ production (in association with $0-$jet and $1-$jet) kept as floating parameters. The $WZ$ production cross section is scaled based on jet multiplicity at particle level. This procedure accounts for possible different effects in the SR and $WZ$ CR of the bin-to-bin migration between the reconstructed and particle level distributions. Contributions of DPS \ssww events with $\tau$-leptons from at least one of the $W$ boson decays, with the $\tau$-lepton decaying leptonically to an electron or a muon, are scaled with the DPS \ssww normalization assuming lepton universality in $W$ decays and correspond to approximately 5\% of the total signal yield in the SR.  The systematic uncertainties are included as nuisance parameters ~\cite{Rolke:2004mj} with Gaussian priors. The nuisance parameters are adjusted in the fit with the shape and normalization of each distribution varying within the specified constraints.

The particle level jets used for the jet multiplicity normalization of the $WZ$ process are reconstructed from stable particles with a lifetime of $\tau>30$ ps in the simulation after parton showering, hadronization, and decay of particles with $\tau < 30$ ps. The \antikt algorithm with a radius parameter of \(R=0.4\) is used where muons, electrons, neutrinos, and photons associated with the $W$ and $Z$ boson decays excluded. The particle level jets with $|\eta|<4.5$ and $\pT > 30~\GeV$ are considered. The small contribution of the $WZ$ production in association with two or more jets is normalized to the SM prediction and allowed to vary within the uncertainty. The simulated $WZ$ \SHERPA samples are used to account for bin-to-bin migration effects between the reconstructed and particle level distributions.

\begin{figure}[t!bp]
\centering
\subfigure[]{\includegraphics[width=.95\textwidth]{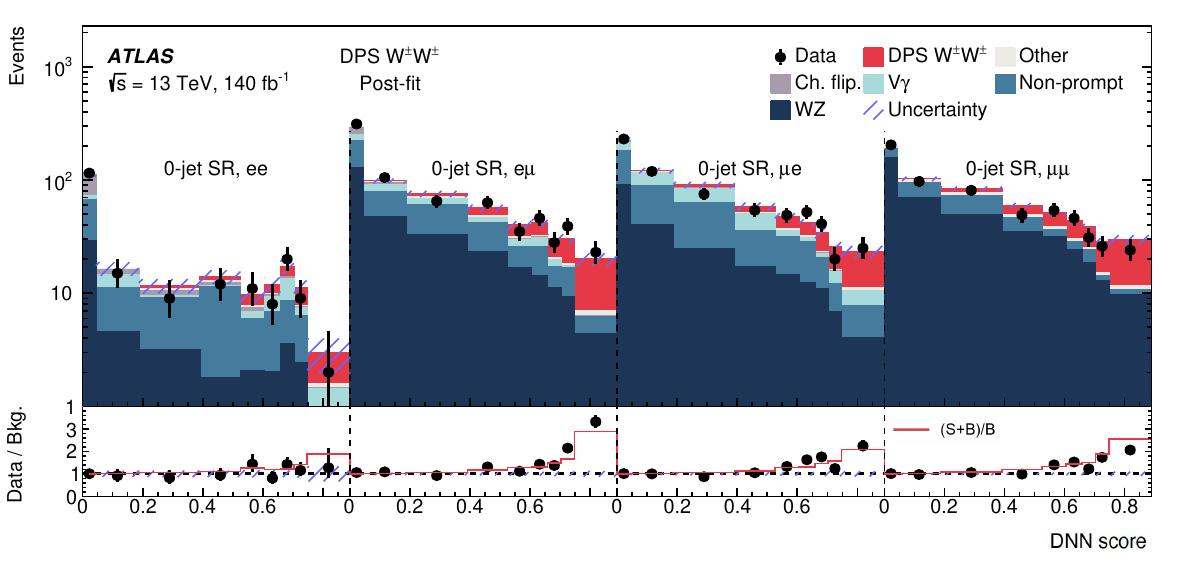}}
\subfigure[]{\includegraphics[width=.95\textwidth]{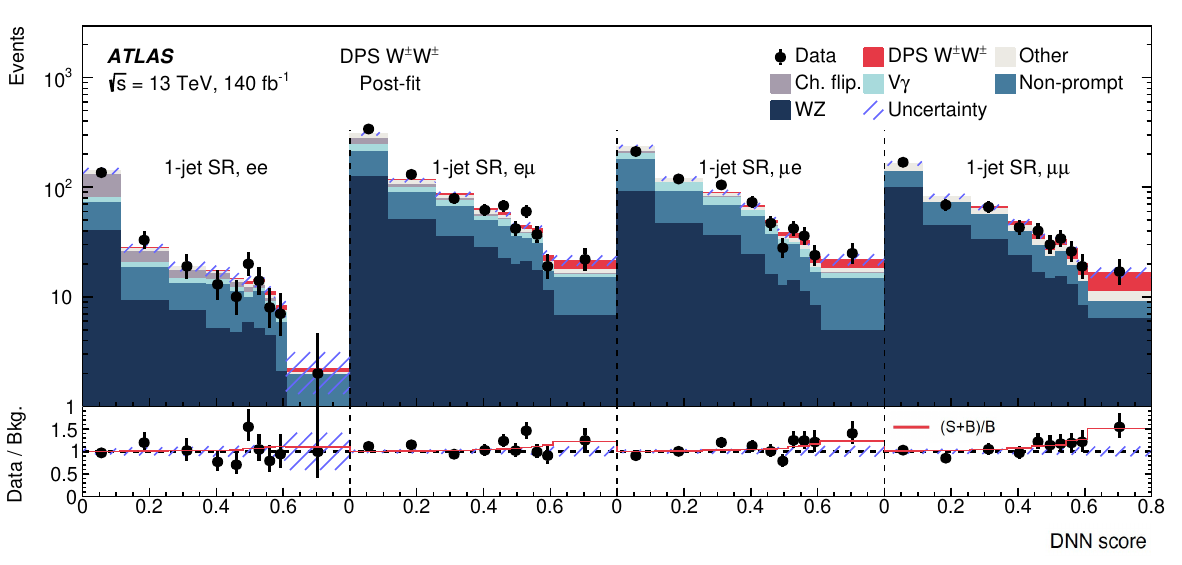}}
\caption{Distributions of the DNN scores in the SR for each four dilepton final states in the (a) $0-$jet and (b) $1-$jet categories. The predicted yields are shown with their best-fit normalization and shape. The shaded area surrounding the expectation represents the total uncertainties in the predicted yields. The ratios of the observed yields to the total background predictions are shown by the points in the bottom panels. The solid line in the bottom panel represents the signal-to-background ratio (1+S/B). The shaded area in the bottom panels represents the relative uncertainties in the predicted yields. The ``Other'' category in the legend combines SPS \ssww, $ZZ$, $VVV$, top quark, and pileup background processes.}
\label{fig:sr_plots}
\end{figure}

\begin{figure}[t!bp]
\centering
\subfigure[]{\includegraphics[width=0.45\linewidth]{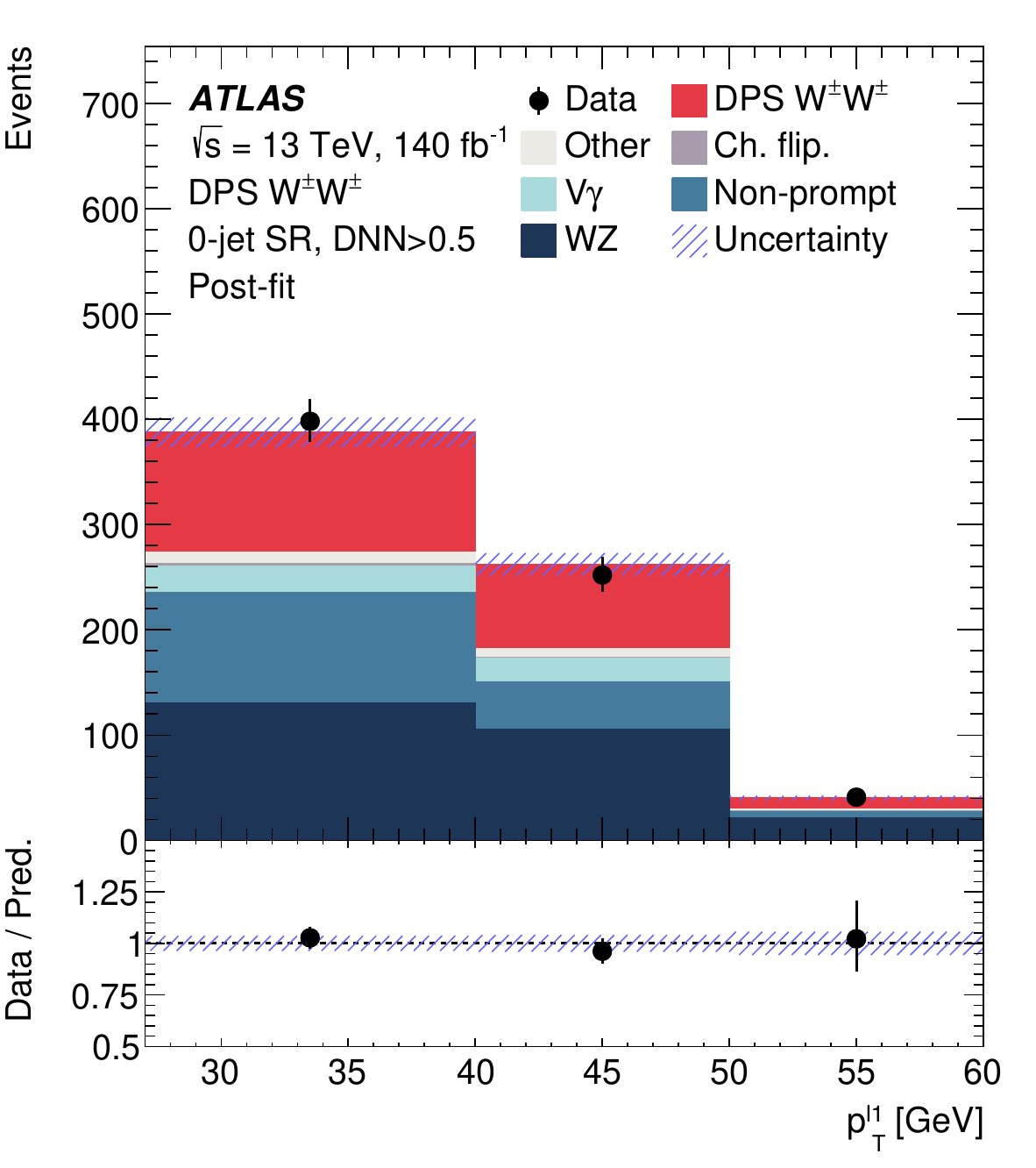}}
\subfigure[]{\includegraphics[width=0.45\linewidth]{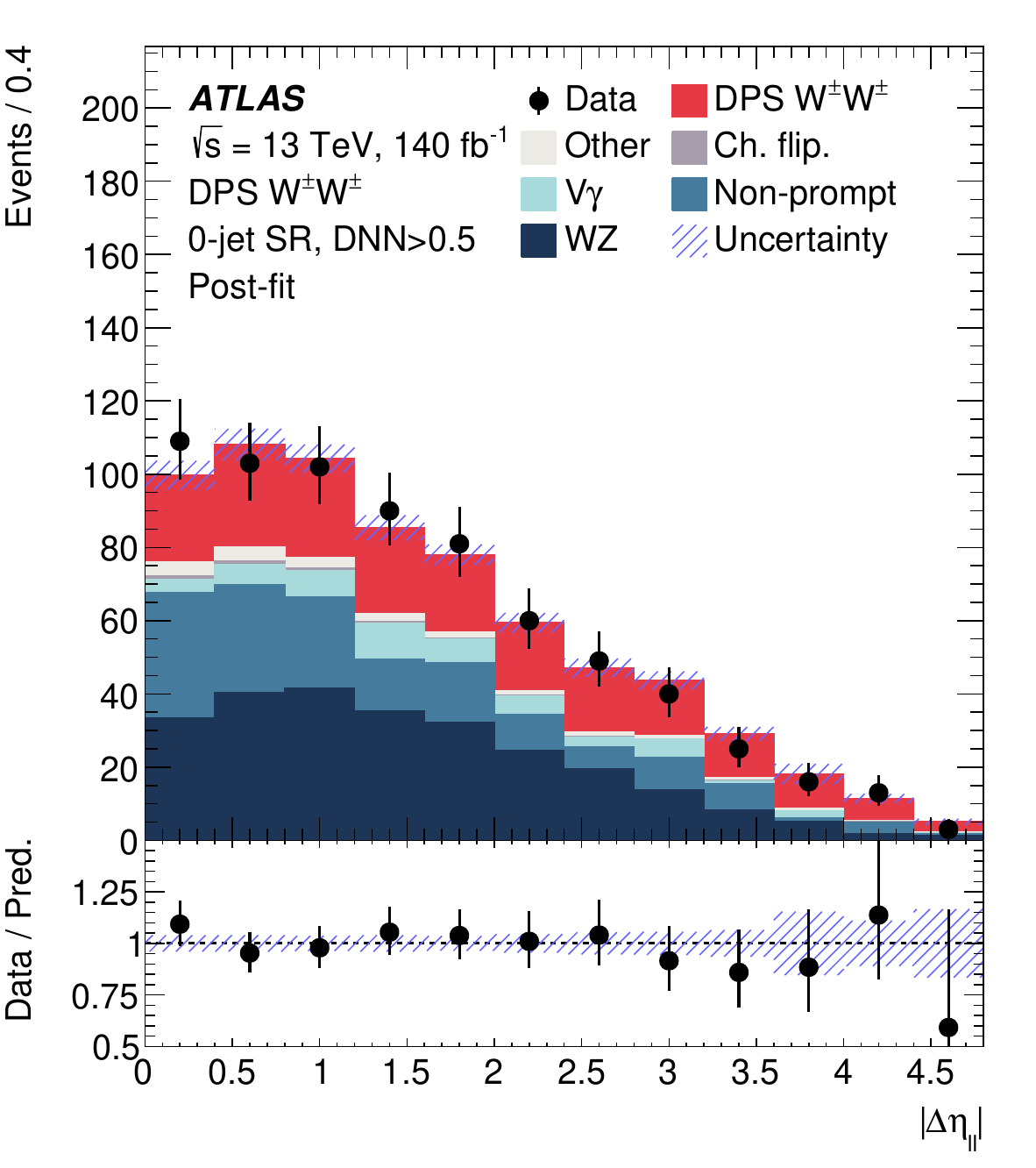}}
\subfigure[]{\includegraphics[width=0.45\linewidth]{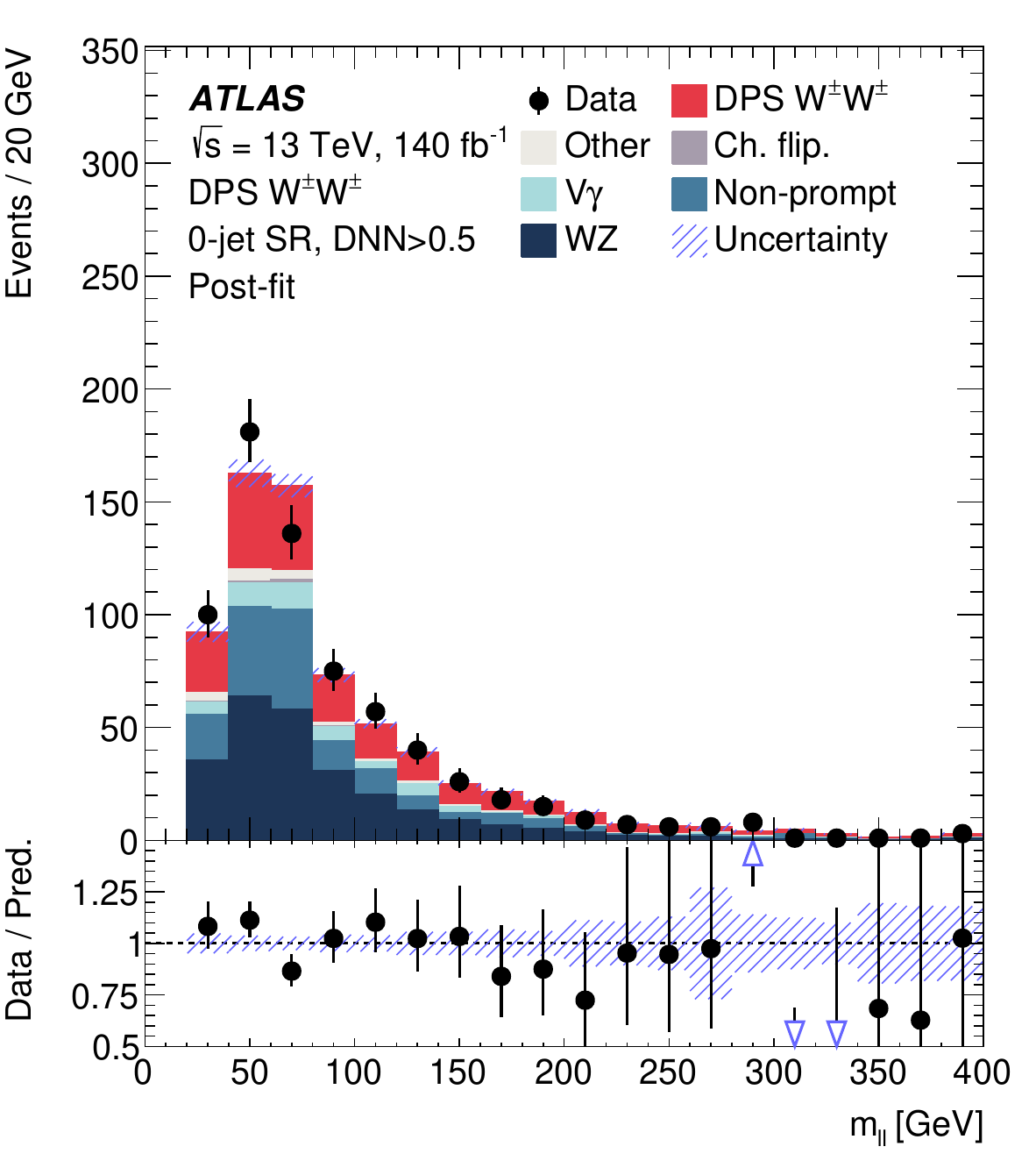}}
\subfigure[]{\includegraphics[width=0.45\linewidth]{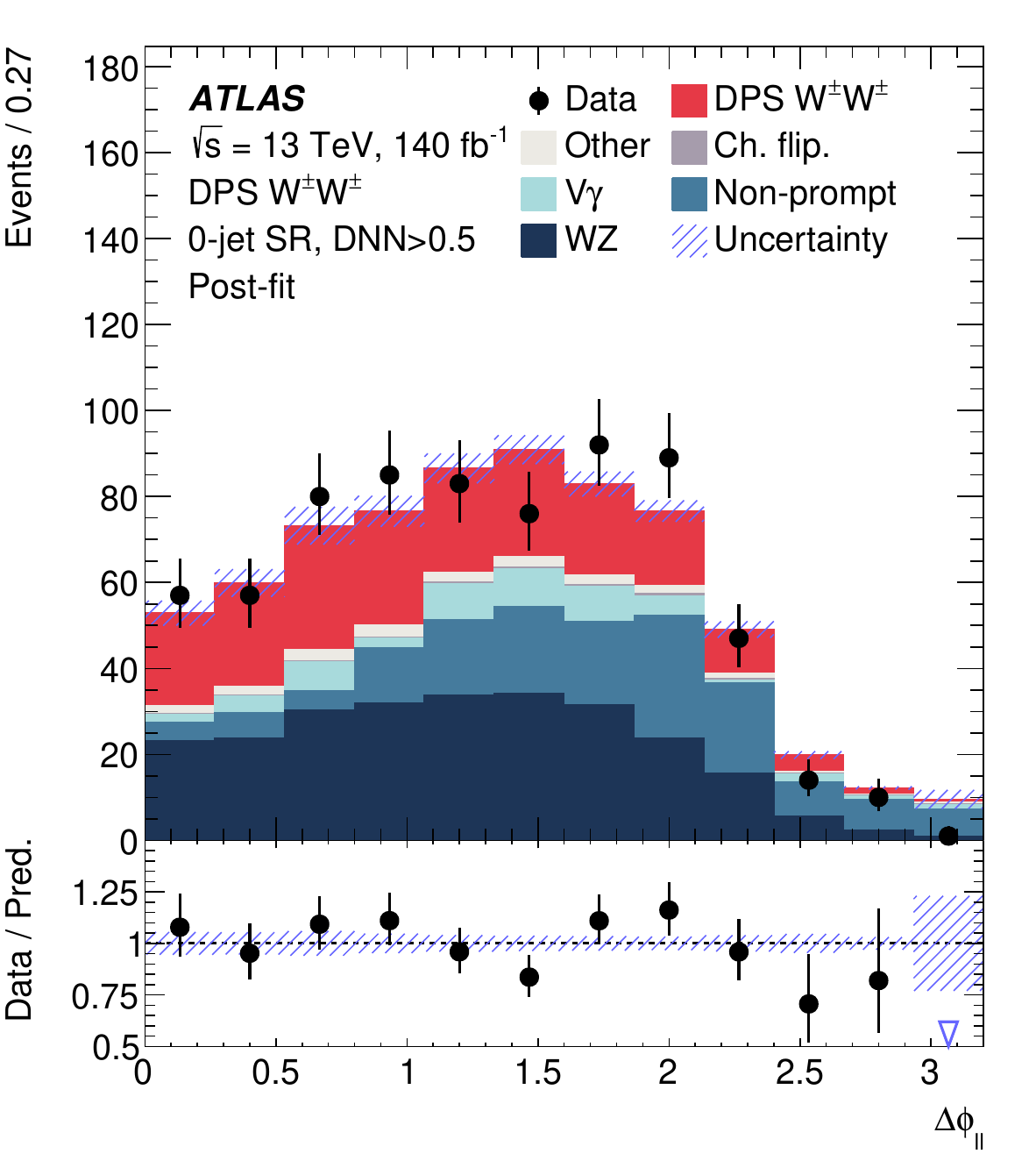}}
\caption{Post-fit distributions of (a) $\pt^{\ell 1}$, (b) $|\Delta \eta_{\ell\ell}|$, (c)  $m_{\ell \ell}$, and (d) $\Delta \phi_{\ell\ell}$ in the $0-$jet category in a region with the signal DNN score greater than 0.5. The distributions for the combined four dilepton final states are shown.  The predicted yields are shown with their best-fit normalization and shape. The shaded area around the expectation represents the total uncertainty in the predicted yields. The ratios of the observed yields to the total predictions are shown by the points in the bottom panels. The ``Other'' category in the legend combines SPS \ssww, $ZZ$, $VVV$, top quark, and pileup background processes.}
\label{fig:postfit_0jetSR_lepvar}
\end{figure}

The signal DNN score distributions in each category are shown in Figure~\ref{fig:sr_plots} with the normalizations and nuisance parameters adjusted by the fit. A good separation between signal and background is achieved. The post-fit yields in the $0-$jet and $1-$jet categories are shown in Tables~\ref{tab:SR0j_yields} and~\ref{tab:SR1j_yields}, respectively. The normalization factors for the $WZ$ production in association with $0-$jet and $1-$jet are $0.96 \pm 0.05$ and $0.94 \pm 0.05$, respectively, consistent with the values reported in Ref.~\cite{STDM-2018-03}. No uncertainties are significantly constrained or pulled in the simultaneous fit. The post-fit distributions  of the $m_{\ell \ell}$, $|\Delta \eta_{\ell\ell}|$, $\Delta \phi_{\ell\ell}$, and $\pt^{\ell 1}$ in the $0-$jet SR category in a region with the signal DNN score greater than 0.5 are shown in Figure~\ref{fig:postfit_0jetSR_lepvar}.

\begin{table}[t!bp]
\caption{Post-fit expected signal and background yields and observed data events in the SR $0-$jet category. The yields are shown for the four dilepton final states. The total uncertainties in the predicted yields are shown.  The uncertainties in the ``Combined" result take into account correlations of systematic uncertainties across the final states The ``Other'' category in the legend combines SPS \ssww, $ZZ$, $VVV$, top quark, and pileup background processes.}
\label{tab:SR0j_yields}
\centering
\begin{tabular}{
l
r@{$\,\pm\,$}l
r@{$\,\pm\,$}l
r@{$\,\pm\,$}l
r@{$\,\pm\,$}l
r@{$\,\pm\,$}l
}
\toprule
Process & \multicolumn{2}{c}{$\hspace{2mm}ee$} & \multicolumn{2}{c}{\hspace{1mm}$e\mu$} & \multicolumn{2}{c}{\hspace{2mm}$\mu e$} & \multicolumn{2}{c}{\hspace{3mm}$\mu\mu$} & \multicolumn{2}{c}{Combined} \\
\midrule
DPS \ssww     &     13.7 & 2.0 & 68 & 9 & 66 & 9 & 90 & 12 & 237 & 32 \\
$WZ$              &     48.9 & 2.9 & 288 & 12 & 221 & 9 & 409 & 13 & 968 & 31 \\
Non-prompt		&     78 & 10 & 204 & 15 & 248 & 18 & 108 & 14 & 640 & 50 \\
Charge flip		&     43 & 10 & 39 & 6 & 13.0 & 1.9 & \multicolumn{2}{c}{\hspace{2mm}---} & 95 & 17 \\
$V\gamma$               &     20 & 5 & 70 & 15 & 107 & 23 & \multicolumn{2}{c}{\hspace{2mm}---} & 200 & 40 \\
Other			&     5.6 & 0.8 & 18.7 & 2.5 & 14.8 & 1.8 & 31 & 5 & 71 & 9 \\
\midrule
Total                   &     209 & 10 & 688 & 17 & 670 & 19 & 639 & 17 & 2210 & 40 \\
\midrule
Data                  &    \multicolumn{2}{c}{\hspace{2mm}201}  &   \multicolumn{2}{c}{\hspace{3mm}716}  &   \multicolumn{2}{c}{\hspace{2mm}665}  &   \multicolumn{2}{c}{\hspace{2mm}612}  &   \multicolumn{2}{c}{\hspace{3mm}2194}\\
\bottomrule
\end{tabular}
\end{table}

\begin{table}
\caption{Post-fit expected signal and background yields and observed data events in the SR $1-$jet category. The yields are shown for the four dilepton final states. The total uncertainties in the predicted yields are shown.  The uncertainties in the ``Combined" result take into account correlations of systematic uncertainties across the final states. The ``Other'' category combines SPS \ssww, $ZZ$, $VVV$, top quark, and pileup background processes.}
\label{tab:SR1j_yields}
\centering
\begin{tabular}{
l
r@{$\,\pm\,$}l
r@{$\,\pm\,$}l
r@{$\,\pm\,$}l
r@{$\,\pm\,$}l
r@{$\,\pm\,$}l
}
\toprule
Process & \multicolumn{2}{c}{$\hspace{2mm}ee$} & \multicolumn{2}{c}{\hspace{1mm}$e\mu$} & \multicolumn{2}{c}{\hspace{2mm}$\mu e$} & \multicolumn{2}{c}{\hspace{3mm}$\mu\mu$} & \multicolumn{2}{c}{Combined} \\
\midrule
DPS \ssww    &     4.8 & 0.8 & 26 & 4 & 25 & 4 & 35 & 5 & 91 & 14 \\
$WZ$              &     84 & 4 & 337 & 12 & 265 & 10 & 275 & 11 & 959 & 33 \\
Non-prompt              &     79 & 10 & 257 & 19 & 271 & 19 & 152 & 19 & 760 & 60 \\
Charge flip             &     65 & 15 & 49 & 7 & 14.5 & 2.1 & \multicolumn{2}{c}{\hspace{2mm}---} & 128 & 24 \\
$V\gamma$               &     18 & 4 & 73 & 17 & 84 & 18 & \multicolumn{2}{c}{\hspace{2mm}---} & 170 & 40 \\
Other                   &     17.8 & 1.4 & 63 & 5 & 54 & 4 & 64 & 4 & 199 & 14 \\
\midrule
Total                   &     269 & 13 & 806 & 18 & 714 & 18 & 525 & 17 & 2310 & 40 \\
\midrule
Data                  &    \multicolumn{2}{c}{\hspace{2mm}262}  &   \multicolumn{2}{c}{\hspace{2mm}860}  &   \multicolumn{2}{c}{\hspace{2mm}711}  &   \multicolumn{2}{c}{\hspace{2mm}513}  &   \multicolumn{2}{c}{\hspace{3mm}2346}\\
\bottomrule
\end{tabular}
\end{table}

The visible excess in the DNN score distributions is quantified by calculating the corresponding $p$--value using a profile likelihood-ratio test statistic in the asymptotic approximation~\cite{Cowan:2010js}. The background-only hypothesis is rejected with an observed significance of 8.8 standard deviations. The fiducial cross section times branching fraction for the DPS $\ssww$ process ($\sigma \mathcal{B} (\ssww)$) is extracted in the fiducial region defined to be as close as possible to the data event selection requirements. The fiducial phase space definition requires two prompt leptons at particle level ($e$ or $\mu$) with $\pT > 27~\GeV$ and $|\eta| < 2.5$, dressed by adding the four-momenta of nearby prompt photons within a small cone of $\Delta R<0.1$, with $m_{\ell \ell}$ > 20~\GeV. Contributions of events with $\tau$-leptons from at least one of the $W$ boson decays are not included in the fiducial region definition. The number of particle level jets, as defined above, is required to be less than 2. The electrons in the dielectron final state are required to have $|\eta|$ less than 1.37. Events with the dielectron invariant mass in the range $|m_{ee}-m_Z| < 15~\GeV$ are rejected. The particle-level \MET is reconstructed from the visible final-state objects and must be greater than 30~\GeV. The fraction of signal events generated outside the fiducial region but satisfying the SR selection at detector level is approximately 14\%. The measured DPS $\sigma \mathcal{B} (\ssww)$ is $4.59 \pm 0.64$ fb. This value cannot be directly compared to the corresponding DPS $\sigma \mathcal{B} (\ssww)$ CMS measurement~\cite{CMS-SMP-21-013} as the definitions of the fiducial regions differ. A summary of the fractional uncertainties in the DPS $\sigma \mathcal{B}$ (\ssww) measurement is shown in Table~\ref{tab:tot_uncert}. The results are driven by the statistical uncertainty of the data. Excluding the $|\eta|<1.37$ requirement for the dielectron final state in the fiducial region definition yields a DPS $\sigma \mathcal{B} (\ssww)$ of $5.52 \pm 0.77$ fb. The DPS $\sigma \mathcal{B} (\ssww)$ in this fiducial region is approximately 20\% larger than in the nominal fiducial region, as the charged leptons tend to have large $|\eta|$ values.

\begin{table}
\caption{The fractional uncertainty of different components in the DPS $\sigma \mathcal{B} (\ssww)$ measurement. The contribution of a systematic uncertainty (uncertainty group) to the total uncertainty is evaluated by fixing the respective nuisance parameter(s) to its (their) best-fit value(s), redoing the fit, and subtracting the uncertainties in the cross section in quadrature. The procedure is implemented incrementally such that the sum in quadrature of the grouped systematic and statistical uncertainties corresponds to the total cross section uncertainty by construction. Lepton uncertainties encompass the effects of the calibration of lepton energy or momentum scale and resolution, as well as the lepton trigger, reconstruction, identification, and isolation efficiencies. The ``DPS \ssww model'' uncertainty is related to the differences in the shapes of the signal distributions predicted by the \PYTHIA and \HERWIG samples. The ``Model statistical'' is related to the effect of a finite number of data events used for data-driven background estimates and of MC events.}
\label{tab:tot_uncert}
\centering
\begin{tabular}{p{2ex}lcc} %
\toprule %
\multicolumn{2}{l}{Source} & Uncertainty [\%] \\ %
\midrule %
\multicolumn{2}{l}{Experimental} & 4.7\\ %
\midrule %
& Electrons & ${\hphantom{0}0.4}$ \\ %
& Muons & ${\hphantom{0}0.8}$ \\ %
& Jets & ${\hphantom{0}3.1}$ \\ %
& $E_{\mathrm{T}}^{\mathrm{miss}}$ & ${\hphantom{0}1.5}$ \\ %
& Flavor tagging & ${\hphantom{0}0.3}$ \\ %
& Non-prompt leptons & ${\hphantom{0}2.6}$ \\ %
& Charge flip & ${\hphantom{0}0.6}$ \\ %
& Pileup reweighting & ${\hphantom{0}1.3}$ \\ %
& Luminosity & ${\hphantom{0}0.8}$ \\ %
\midrule %
\multicolumn{2}{l}{Modeling} & 1.5\\ %
\midrule %
& DPS \ssww scale & ${\hphantom{0}0.2}$ \\ %
& DPS \ssww model & ${\hphantom{0}0.3}$ \\ %
& SPS \ssww scale, PDF \& $\alpha_S$ & ${\hphantom{0}0.2}$ \\ %
& $WZ$ scale, PDF \& $\alpha_S$ & ${\hphantom{0}0.8}$ \\ %
& $WZ$ normalization & ${\hphantom{0}0.1}$ \\ %
& Other background normalizations  & ${\hphantom{0}1.1}$ \\ %
& Model statistical & ${\hphantom{0}0.6}$ \\ %
\midrule %
\multicolumn{2}{l}{ Experimental and modeling} & ${\hphantom{0}5.0}$ \\ %
\multicolumn{2}{l}{ Data statistical} & ${\hphantom{0}13}$ \\ %
\midrule %
\multicolumn{2}{l}{ Total} & ${\hphantom{0}14}$ \\ %
\bottomrule %
\end{tabular} %
\end{table}

\begin{figure}[t!bp]
\centering
\includegraphics[width=0.80\linewidth]{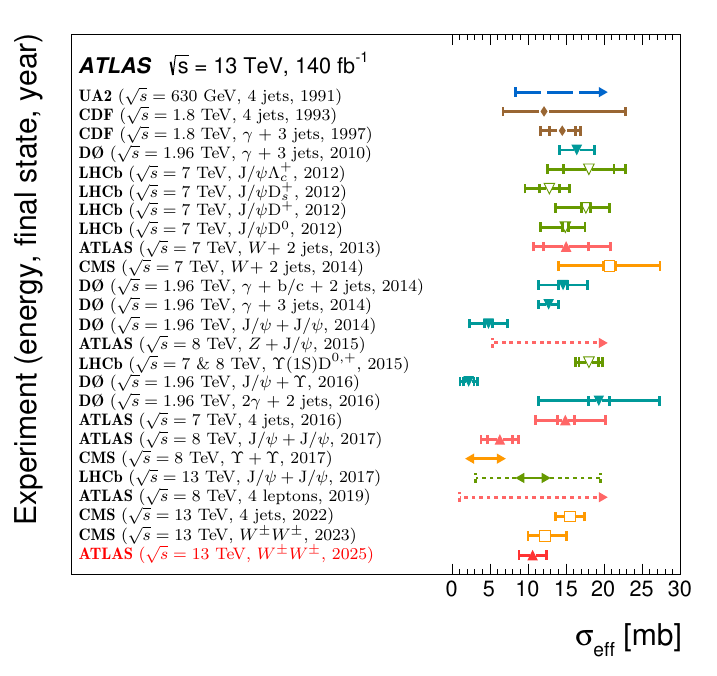}
\caption{Summary of measurements and limits on $\sigma_{\mathrm{eff}}$, determined in diﬀerent experiments~\cite{STDM-2011-29,CMS-FSQ-12-028,sigeff4,CMS-GEN-14-001,sigeff3,BPHY-2014-01,Aaij:2015wpa,Abazov:2015fbl,CMS-BPH-21-004,sigeff5,sigeff6,cdf:dps,CDF:dps2,LHCb:dps,CMS-FSQ-16-005,STDM-2017-15,CMS-SMP-20-007,CMS-SMP-20-009,D0:2014vql,D0:2014owy, D0:2015rpo, STDM-2015-08, BPHY-2015-02, CMS-BPH-14-008, LHCb:2016wuo}, sorted chronologically. The measurements that were made by different experiments are denoted by different symbols and colors. The inner error bars represent statistical uncertainties and the outer error bars correspond to the total uncertainty. Dashed arrows indicate lower limits. Lines with arrows on both ends represent ranges of the $\sigma_{\mathrm{eff}}$ values, determined within a single publication. In the case of the double $J/\psi$ measurement by LHCb~\cite{LHCb:2016wuo}, the dashed line denotes the upper and lower uncertainties. The AFS measurement of $\sigma_{\mathrm{eff}}= 5$\,mb at $\sqrt{s}=63~\GeV$~\cite{sigeff5} was published without uncertainties and is not included in the plot.}
\label{fig:summary}
\end{figure}

The DPS $\sigma \mathcal{B}$ (\ssww) measurement is used to extract the value of $\sigma_{\mathrm{eff}}$ using Eq.~\ref{eq:DPS}. A common fiducial region for the cross sections of the $W^{+}$, $W^{-}$, and DPS \ssww processes needs to be defined for the correct determination of $\sigma_{\mathrm{eff}}$. Therefore, the measured $\sigma \mathcal{B}$ (\ssww) value is extrapolated to a region having the same requirements as the analysis fiducial region but removing the particle-level requirements on $m_{\ell \ell}$ and \MET, and the requirements on electron $|\eta| < 1.37$ and $|m_{ee}-m_Z| < 15~\GeV$ in the dielectron final state. The extrapolation is performed using the $\PYTHIA$ signal samples. The extrapolation factor obtained with $\PYTHIA$ is $1.75 \pm 0.04$, where the uncertainty includes the QCD scale variations. The extrapolation factor obtained with the $\HERWIG$ sample is in excellent agreement with this value. The cross sections of the $W^{+}$ and $W^{-}$ processes in this common fiducial region are calculated using the \SHERPA[2.2.11] event generator at NLO in QCD. The predicted cross sections times the per-flavor leptonic branching fractions are \(\sigma_{W^{+}} = 5.23 \pm 0.34\) nb and \(\sigma_{W^{-}} = 4.00 \pm 0.22\) nb where the uncertainties include the QCD scale variations and are propagated in the determination of $\sigma_{\mathrm{eff}}$. The contributions of events with two SPS $W+1-$jet processes are not included in the numerator of Eq.~\ref{eq:DPS} as the number of particle level jets is required to be less than 2 for the DPS $\sigma \mathcal{B} (\ssww)$. The resulting value of $\sigma_{\mathrm{eff}}$ is $10.6 \pm 1.8$ mb consistent with the corresponding CMS measurement of $12.2^{+2.9}_{-2.2}$ mb~\cite{CMS-SMP-21-013} and with previous determinations at hadron colliders as shown in Figure~\ref{fig:summary}.

\FloatBarrier

\section{Conclusion}

This letter reports the measurement of double parton scattering in same-sign $W$ boson pair production with the ATLAS detector. The dataset used corresponds to an integrated luminosity of 140~fb$^{-1}$ of proton--proton collisions at a center-of-mass energy of 13~TeV, collected during Run 2 of the Large Hadron Collider. The study is performed in final states including two same-charge leptons, electron or muon, missing transverse momentum, and up to one jet. An excess of events over the expected background contributions is reported with a significance of 8.8 standard deviations. The measured fiducial cross section times leptonic branching fraction is $4.59 \pm 0.64$ fb. The measurement corresponds to a double parton scattering effective cross section of $10.6 \pm 1.8$ mb.

\section*{Acknowledgements}
%

%

% The next lines are included from the .//acknowledgements/Acknowledgements.tex input file
%
%

%
%

We thank CERN for the very successful operation of the LHC and its injectors, as well as the support staff at
CERN and at our institutions worldwide without whom ATLAS could not be operated efficiently.

The crucial computing support from all WLCG partners is acknowledged gratefully, in particular from CERN, the ATLAS Tier-1 facilities at TRIUMF/SFU (Canada), NDGF (Denmark, Norway, Sweden), CC-IN2P3 (France), KIT/GridKA (Germany), INFN-CNAF (Italy), NL-T1 (Netherlands), PIC (Spain), RAL (UK) and BNL (USA), the Tier-2 facilities worldwide and large non-WLCG resource providers. Major contributors of computing resources are listed in Ref.~\cite{ATL-SOFT-PUB-2025-001}.

We gratefully acknowledge the support of ANPCyT, Argentina; YerPhI, Armenia; ARC, Australia; BMWFW and FWF, Austria; ANAS, Azerbaijan; CNPq and FAPESP, Brazil; NSERC, NRC and CFI, Canada; CERN; ANID, Chile; CAS, MOST and NSFC, China; Minciencias, Colombia; MEYS CR, Czech Republic; DNRF and DNSRC, Denmark; IN2P3-CNRS and CEA-DRF/IRFU, France; SRNSFG, Georgia; BMFTR, HGF and MPG, Germany; GSRI, Greece; RGC and Hong Kong SAR, China; ICHEP and Academy of Sciences and Humanities, Israel; INFN, Italy; MEXT and JSPS, Japan; CNRST, Morocco; NWO, Netherlands; RCN, Norway; MNiSW, Poland; FCT, Portugal; MNE/IFA, Romania; MSTDI, Serbia; MSSR, Slovakia; ARIS and MVZI, Slovenia; DSI/NRF, South Africa; MICIU/AEI, Spain; SRC and Wallenberg Foundation, Sweden; SERI, SNSF and Cantons of Bern and Geneva, Switzerland; NSTC, Taipei; TENMAK, T\"urkiye; STFC/UKRI, United Kingdom; DOE and NSF, United States of America.

Individual groups and members have received support from BCKDF, CANARIE, CRC and DRAC, Canada; CERN-CZ, FORTE and PRIMUS, Czech Republic; COST, ERC, ERDF, Horizon 2020, ICSC-NextGenerationEU and Marie Sk{\l}odowska-Curie Actions, European Union; Investissements d'Avenir Labex, Investissements d'Avenir Idex and ANR, France; DFG and AvH Foundation, Germany; Herakleitos, Thales and Aristeia programmes co-financed by EU-ESF and the Greek NSRF, Greece; BSF-NSF and MINERVA, Israel; NCN and NAWA, Poland; La Caixa Banking Foundation, CERCA Programme Generalitat de Catalunya and PROMETEO and GenT Programmes Generalitat Valenciana, Spain; G\"{o}ran Gustafssons Stiftelse, Sweden; The Royal Society and Leverhulme Trust, United Kingdom.

In addition, individual members wish to acknowledge support from CERN: European Organization for Nuclear Research (CERN DOCT); Chile: Agencia Nacional de Investigaci\'on y Desarrollo (FONDECYT 1230812, FONDECYT 1240864, Fondecyt 3240661); China: Chinese Ministry of Science and Technology (MOST-2023YFA1605700, MOST-2023YFA1609300), National Natural Science Foundation of China (NSFC - 12175119, NSFC 12275265); Czech Republic: Czech Science Foundation (GACR - 24-11373S), Ministry of Education Youth and Sports (ERC-CZ-LL2327, FORTE CZ.02.01.01/00/22\_008/0004632), PRIMUS Research Programme (PRIMUS/21/SCI/017); EU: H2020 European Research Council (ERC - 101002463); European Union: European Research Council (BARD No. 101116429, ERC - 948254, ERC 101089007), European Regional Development Fund (SMASH COFUND 101081355, SLO ERDF), Horizon 2020 Framework Programme (MUCCA - CHIST-ERA-19-XAI-00), European Union, Future Artificial Intelligence Research (FAIR-NextGenerationEU PE00000013), Italian Center for High Performance Computing, Big Data and Quantum Computing (ICSC, NextGenerationEU); France: Agence Nationale de la Recherche (ANR-21-CE31-0022, ANR-22-EDIR-0002); Germany: Baden-Württemberg Stiftung (BW Stiftung-Postdoc Eliteprogramme), Deutsche Forschungsgemeinschaft (DFG - 469666862, DFG - CR 312/5-2); China: Research Grants Council (GRF); Italy: Istituto Nazionale di Fisica Nucleare (ICSC, NextGenerationEU), Ministero dell'Università e della Ricerca (NextGenEU 153D23001490006 M4C2.1.1, NextGenEU I53D23000820006 M4C2.1.1, NextGenEU I53D23001490006 M4C2.1.1, SOE2024\_0000023); Japan: Japan Society for the Promotion of Science (JSPS KAKENHI JP22H01227, JSPS KAKENHI JP22H04944, JSPS KAKENHI JP22KK0227, JSPS KAKENHI JP24K23939, JSPS KAKENHI JP25H00650, JSPS KAKENHI JP25H01291, JSPS KAKENHI JP25K01023); Norway: Research Council of Norway (RCN-314472); Poland: Ministry of Science and Higher Education (IDUB AGH, POB8, D4 no 9722), Polish National Science Centre (NCN 2021/42/E/ST2/00350, NCN OPUS 2023/51/B/ST2/02507, NCN OPUS nr 2022/47/B/ST2/03059, NCN UMO-2019/34/E/ST2/00393, UMO-2022/47/O/ST2/00148, UMO-2023/49/B/ST2/04085, UMO-2023/51/B/ST2/00920, UMO-2024/53/N/ST2/00869); Portugal: Foundation for Science and Technology (FCT); Spain: Ministry of Science and Innovation (MCIN \& NextGenEU PCI2022-135018-2, MICIN \& FEDER PID2021-125273NB, RYC2019-028510-I, RYC2020-030254-I, RYC2021-031273-I, RYC2022-038164-I); Sweden: Carl Trygger Foundation (Carl Trygger Foundation CTS 22:2312), Swedish Research Council (Swedish Research Council 2023-04654, VR 2021-03651, VR 2022-03845, VR 2022-04683, VR 2023-03403, VR 2024-05451), Knut and Alice Wallenberg Foundation (KAW 2018.0458, KAW 2022.0358, KAW 2023.0366); Switzerland: Swiss National Science Foundation (SNSF - PCEFP2\_194658); United Kingdom: Leverhulme Trust (Leverhulme Trust RPG-2020-004), Royal Society (NIF-R1-231091); United States of America: U.S. Department of Energy (ECA DE-AC02-76SF00515), Neubauer Family Foundation.

%
%

% End of text imported from the .//acknowledgements/Acknowledgements.tex input file

%
\clearpage
\printbibliography
\clearpage
% ATLAS Collaboration author list
% Reference date of STDM-2022-18 is 2025-04-14
% Author list last updated on date 19-AUG-25
% Data extracted on 19-Aug-2025 for paper reference STDM-2022-18
% at 11:20am
 
\begin{flushleft}
\hypersetup{urlcolor=black}
{\Large The ATLAS Collaboration}

\bigskip

\AtlasOrcid[0000-0002-6665-4934]{G.~Aad}$^\textrm{\scriptsize 104}$,
\AtlasOrcid[0000-0001-7616-1554]{E.~Aakvaag}$^\textrm{\scriptsize 17}$,
\AtlasOrcid[0000-0002-5888-2734]{B.~Abbott}$^\textrm{\scriptsize 123}$,
\AtlasOrcid[0000-0002-0287-5869]{S.~Abdelhameed}$^\textrm{\scriptsize 119a}$,
\AtlasOrcid[0000-0002-1002-1652]{K.~Abeling}$^\textrm{\scriptsize 55}$,
\AtlasOrcid[0000-0001-5763-2760]{N.J.~Abicht}$^\textrm{\scriptsize 49}$,
\AtlasOrcid[0000-0002-8496-9294]{S.H.~Abidi}$^\textrm{\scriptsize 30}$,
\AtlasOrcid[0009-0003-6578-220X]{M.~Aboelela}$^\textrm{\scriptsize 45}$,
\AtlasOrcid[0000-0002-9987-2292]{A.~Aboulhorma}$^\textrm{\scriptsize 36e}$,
\AtlasOrcid[0000-0001-5329-6640]{H.~Abramowicz}$^\textrm{\scriptsize 157}$
% ...
\bigskip
\\

% ...
$^{17}$Department for Physics and Technology, University of Bergen, Bergen; Norway.\\
% ...
$^{30}$Physics Department, Brookhaven National Laboratory, Upton NY; United States of America.\\
% ...
$^{36}$$^{(a)}$Facult\'e des Sciences Ain Chock, Universit\'e Hassan II de Casablanca;$^{(b)}$Facult\'{e} des Sciences, Universit\'{e} Ibn-Tofail, K\'{e}nitra;$^{(c)}$Facult\'e des Sciences Semlalia, Universit\'e Cadi Ayyad, LPHEA-Marrakech;$^{(d)}$LPMR, Facult\'e des Sciences, Universit\'e Mohamed Premier, Oujda;$^{(e)}$Facult\'e des sciences, Universit\'e Mohammed V, Rabat;$^{(f)}$Institute of Applied Physics, Mohammed VI Polytechnic University, Ben Guerir; Morocco.\\
% ...
$^{45}$Physics Department, Southern Methodist University, Dallas TX; United States of America.\\
% ...
$^{49}$Fakult\"{a}t Physik , Technische Universit{\"a}t Dortmund, Dortmund; Germany.\\
% ...
$^{55}$II. Physikalisches Institut, Georg-August-Universit\"{a}t G\"ottingen, G\"ottingen; Germany.\\
% ...
$^{104}$CPPM, Aix-Marseille Universit\'e, CNRS/IN2P3, Marseille; France.\\
% ...
$^{119}$$^{(a)}$New York University Abu Dhabi, Abu Dhabi;$^{(b)}$United Arab Emirates University, Al Ain; United Arab Emirates.\\
% ...
$^{123}$Homer L. Dodge Department of Physics and Astronomy, University of Oklahoma, Norman OK; United States of America.\\
% ...
$^{157}$Raymond and Beverly Sackler School of Physics and Astronomy, Tel Aviv University, Tel Aviv; Israel.\\
% ...
$^{a}$ Also at Affiliated with an institute formerly covered by a cooperation agreement with CERN.\\
$^{b}$ Also at An-Najah National University, Nablus; Palestine.\\
$^{c}$ Also at Borough of Manhattan Community College, City University of New York, New York NY; United States of America.\\
$^{d}$ Also at Center for Interdisciplinary Research and Innovation (CIRI-AUTH), Thessaloniki; Greece.\\
$^{e}$ Also at Centre of Physics of the Universities of Minho and Porto (CF-UM-UP); Portugal.\\
$^{f}$ Also at CERN, Geneva; Switzerland.\\
% ...

\end{flushleft}

% Created with Glance <Atlas.Glance@cern.ch>

%

%
%

%

%
%
%
%

\end{document}